\documentclass[aps,prb,twocolumn,10pt,superscriptaddress,showpacs,amsmath,amssymb,nofootinbib,longbibliography]{revtex4-1}

\usepackage{graphicx}
\usepackage{hyperref}

\newcommand{\figHIST}{
\begin{figure*}
\includegraphics[width=\textwidth]{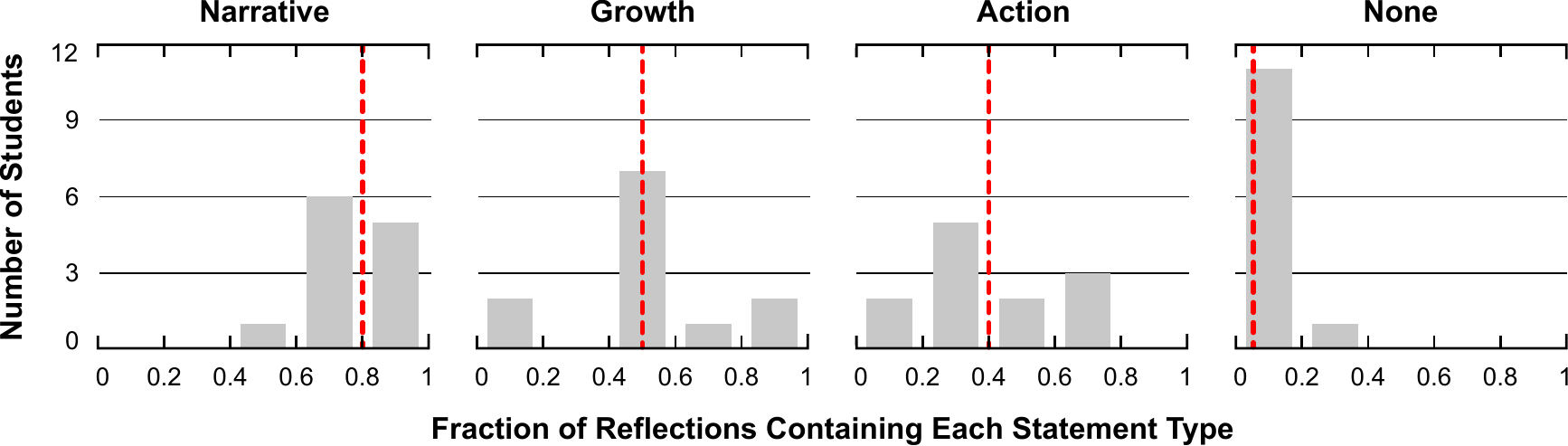}
\caption{\label{fig:hist}Distribution of students for whom a given fraction of reflections contained each type of statement. Statement types include narrative, growth, and action statements, as well as statements that fit into none of these categories. Thick dashed vertical red lines represent the averages of the distributions.}
\end{figure*}
}

\newcommand{\figVENN}{
\begin{figure}
\includegraphics[width=\columnwidth]{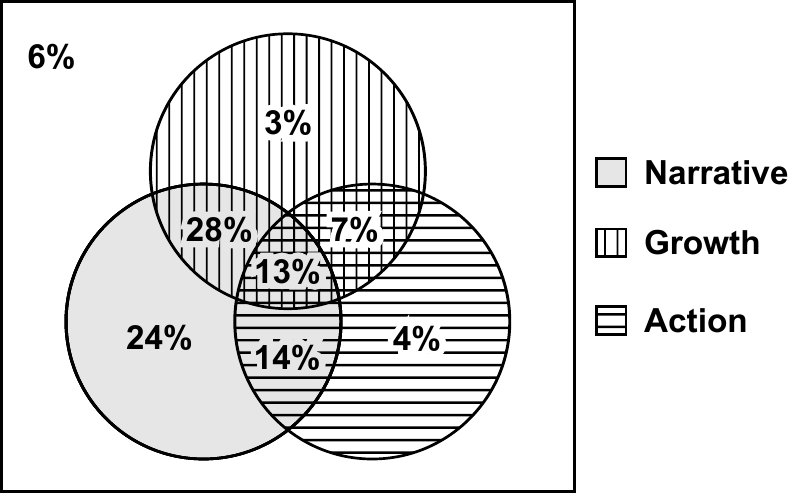}
\caption{\label{fig:venn}Venn diagram showing the presence of narrative, growth, and/or action statements in students completed reflections. Only 6\% of reflections contained none of these statement types, whereas almost two-thirds contained more than one type of statement.}
\end{figure}
}

\newcommand{\figPIE}{
\begin{figure}
\includegraphics[width=\columnwidth]{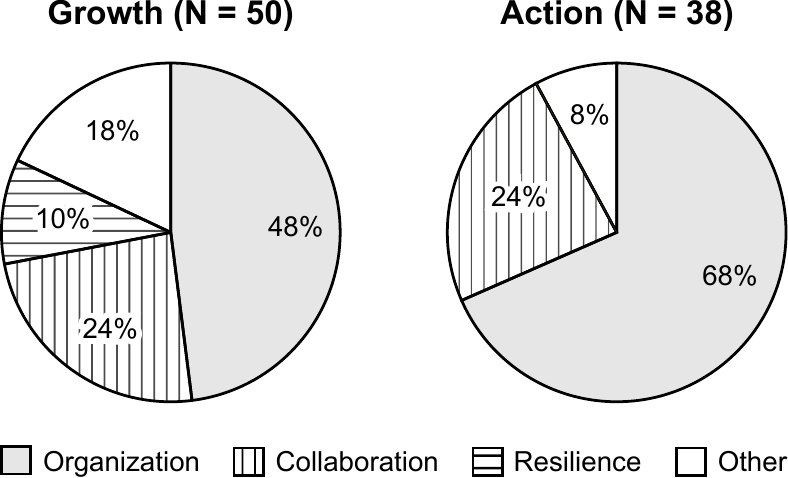}
\caption{\label{fig:pie}Breakdown of growth and action statements by coded skill type. None of the action statements in our dataset concerned making plans to improve resilience.}
\end{figure}
}

\newcommand{\tabIRR}{
\begin{table}
\begin{ruledtabular}
\begin{tabular}{lccl}
Statement Type & Agreement & Kappa & Interpretation \\ \hline
Narrative & 87\% & 0.65 & Substantial \\
Growth & 89\% & 0.77 & Substantial \\
Achievement & 92\% & 0.46 & Moderate \\
Action & 94\% & 0.88 & Almost perfect
\end{tabular}
\end{ruledtabular}
\caption{\label{tab:IRR}Inter-rater reliability metrics for coding of statement types across 71 reflections written by 9 students.}
\end{table}
}

\newcommand{\tabNARR}{
\begin{table}
\begin{ruledtabular}
\begin{tabular}{llcc}
Category & Topic & \multicolumn{2}{c}{Narrative} \\ \cline{1-1}\cline{2-2}\cline{3-4}
{\bf Out-of-class activities} & {\bf All topics} & {\bf 38} & {\bf 49\%} \\
&Weekly homework & 24 & 31\%  \\
&Studying & 9 & 12\%  \\
&Long-term projects & 5 & 6\%   \\ 
{\bf In-class activities} & {\bf All topics} & {\bf 20} & {\bf 26\%} \\
&Working in groups & 12 & 16\%  \\
&Class activities & 5 & 7\%   \\
&Presentations & 3 & 4\%   \\ 
{\bf Other} & {\bf All topics} & {\bf 19} & {\bf 25\%} \\
&Academic & 12 & 16\%   \\
&Non-academic & 7 & 9\%  \\  \hline
{\bf Total} & & {\bf 77} & {\bf 100\%} 
\end{tabular}
\end{ruledtabular}
\caption{\label{tab:narrative-themes}Themes present in students' narrative statements.}
\end{table}
}

\newcommand{\student}[2]{(#1,~Week~#2)}

\begin{document}

\title{Attending to lifelong learning skills through guided reflection in a physics class}
\author{Dimitri R. Dounas-Frazer}
\email{dimitri.dounasfrazer@colorado.edu}
\affiliation{Department of Physics, University of Colorado Boulder,
Boulder, CO 80309, USA}

\author{Daniel L. Reinholz}
\email{daniel.reinholz@colorado.edu}
\affiliation{Center for STEM Learning, University of Colorado Boulder,
Boulder, CO 80309, USA}

\date{\today}

\begin{abstract}
This paper describes a tool, the Guided Reflection Form (GRF), which was used to promote reflection in a modeling-based physics course. Each week, students completed a guided reflection and received feedback from their instructors. These activities were intended to help students become better at the process of reflection, developing skills that they could apply in their future learning. We analyzed student reflections: (1) to provide insight into the reflection process itself and (2) to describe common themes in student reflections. Most students were able to use the GRF to reflect on their learning in meaningful ways. Moreover, the themes present in student reflections provide insights into struggles commonly faced by physics students. We discuss the design of the GRF in detail, so that others may use it as a tool to support student reflections.
\end{abstract}

\maketitle

\section{Introduction}

Learning physics requires more than just understanding physics content; it involves the mastery of scientific practices and learning skills~\cite{national_research_council_adapting_2013}. Many of these learning skills are associated with reflection and self-regulation: the ability to set goals, to make plans to achieve those goals, and to monitor progress in implementing one's plans. These reflective skills are widely recognized as a hallmark of expertise across disciplines~\cite{zimmerman_becoming_2002}. To help students reflect, we developed a tool called the Guided Reflection Form (GRF) through iterative cycles of design, implementation, and analysis~\cite{cobb_design_2003, penuel_organizing_2011}. In addition to focusing on the design and implementation of the GRF, this paper aims to address two research questions: (1) how did the GRF support student reflections and (2) what were the specific areas of focus for student reflections? The context of our study was a \emph{Modeling Instruction}~\cite{Brewe2008} physics course for future physics instructors.

In physics, student reflection has been studied in various contexts, typically connected to students' development of problem-solving skills~\cite{Mason2010}, content knowledge~\cite{Scott2007}, and conceptual understanding~\cite{May2002}. Similarly, one major goal of the GRF is to facilitate development of specific ``lifelong learning" skills--including organization, collaboration, and persistence--because prior work showed that students struggled with these skills~\cite{gandhi_teaching}. More broadly, the GRF was designed to make implicit aspects of learning explicit to students.

Although these lifelong learning skills are implicitly valued by reform physics approaches, such as model-based approaches, they are rarely made explicit in physics courses. Rather, these skills are covered by a ``hidden curriculum'' that students are expected to master to be successful~\cite{jackson_life_1968}. However, Reid and Moore recommend explicit attention to development of time management skills--a subset of organizational skills more generally--in order to improve outcomes for post-secondary students whose parents didn't go to college~\cite{Reid2008}. Similarly, Cohen and Lotan argue that productive collaboration among primary students is a key aspect of achieving equity in heterogeneous classrooms~\cite{Cohen1997}. Hence, we believe that lifelong learning skills must be made explicit to students as a matter of equity.

While the GRF focused reflections on specific skills, the overall aim was to help students develop the ability to reflect on their learning for sustained growth beyond the scope of this single course. In this sense, we were more interested in whether students could learn to become better at reflecting, rather than simply better at collaborating or managing their time. Learning to reflect is nontrivial for students; simply asking students to reflect may not lead to meaningful reflection at all~\cite{anson_reflection_1997} and no single strategy is best for engaging all students in reflection~\cite{Spalding2002}.

Reflection is a highly-valued practice, and further research into physics students' specific areas of reflection will continue to illuminate how this practice can support students in learning physics~\cite{ward_role_2014}. Our analyses of student reflections provide insights into the types of struggles that students faced on a regular basis. The present work differs from previous efforts~\cite{Mason2010,Scott2007,May2002} in that, because our aim is not to further demonstrate the utility of reflection, we do not attempt to connect the practice of reflection to gains in student learning. We focus on the structure and content of student reflections, investigating the effectiveness of the GRF at eliciting reflections that connect students' challenges to concrete plans for growth and improvement. Although our results are limited to the specifics of a single population, we contrast these reflections with prior work in other contexts~\cite{gandhi_teaching, zaniewski_worries} to provide a broader picture of the issues that physics students face. We believe that these reflections provide insight for supporting students in a variety of ways, and should be of broad utility to the physics education community.

\section{Theoretical Framing}

Although researchers have conceptualized reflection in a variety of ways, there is general agreement that the emphasis on reflection in education can be traced back to Dewey~\cite{dewey_how_1933}. We build on this line of work, which focuses on processing \textit{experiences}~\cite{boud_promoting_1996,boyd_reflective_1983,kolb_experiential_1984}. Accordingly, we define reflection as: the act of processing an experience to gain further insight into the experience and better inform future action. This contrasts some more general notions of reflection, that focus on individuals ``thinking deeply''~\cite{moon_learning_1999}. 

We draw from a model of reflection that emphasizes: (1) returning to an experience, (2) attending to feelings and re-evaluating the experience, and (3) linking this processing to action~\cite{boud_promoting_1996}. Returning to the experience involves recalling the important aspects of the experience or recounting them to others. Bringing the experience into consciousness allows an individual to make active and aware decisions about their learning; if the experience remains unconscious, this type of deeper processing is difficult to achieve. Moreover, when the experience is brought back up into one's consciousness, it is often possible for one to evaluate the experience from a more distanced, objective perspective.

Attending to feelings involves focusing on positive feelings (e.g., possible benefits from processing events) and removing obstructing feelings (e.g., removing impediments to future success). By attending explicitly to one's feelings related to an event, it becomes possible to understand the role of those feelings in the how the event was initially interpreted and experienced. It also allows for explicit cognitive processing of the feelings (e.g., by consciously recognizing that one is afraid of failing at a particular type of task, it becomes more possible to address that fear). This processing leads to the development of new goals and understandings.

Once an experience has been processed and re-evaluated, ideally it should lead to a new course of action; this is the ultimate purpose of reflection. Unfortunately, even if individuals are able to generate a course of action, they will not necessarily be able to implement it effectively~\cite{argyris_theories_1976}. As a matter design, this suggests that careful attention must be paid to plans that are created and whether or not they are implemented.

Building on this framework~\cite{boud_promoting_1996}, we conceptualized effective reflection as requiring a learner to answer three questions:
\begin{enumerate}
\item What experience would you like to improve upon?
\item What is your goal for improvement?
\item What is your plan to reach your goal?
\end{enumerate}
These three questions can be seen as relating to each of the areas discussed above~\cite{boud_promoting_1996}.

This focus on goals is aligned well with literature on self-regulated learning~\cite{zimmerman_becoming_2002}. Self-regulation refers to ``self-generated thoughts, feelings, and behaviors that are oriented to attaining goals'' (Zimmerman, 2002, p. 65). In other words, self-regulation focuses on one's ability to set goals, adopt strategies for meeting the goals, and monitor progress towards those goals. These aspects of goal setting and monitoring are embodied in the three guiding questions above.

\section{Design}

Methodologically, we operate within the design-based research paradigm. Design-based research aims to contribute simultaneously to theory and practice through iterative cycles of design, implementation, and analysis~\cite{cobb_design_2003}. This methodology emerged to address the need to develop practical interventions in real-life contexts~\cite{brown_design_1992}. Recently, researchers have argued that practical relevance is an important criterion for the value and rigor of research~\cite{burkhardt_improving_2003,gutierrez_relevance_2014}. 

This paper focuses on the fourth iteration of the GRF. Across these four iterations, the basic structure and goals of the design have remained constant. Students are asked to regularly reflect on their learning processes (generally once a week), and they receive feedback from their instructors about their reflections. Students are asked to reflect on a number of different learning skills and how they apply to their experiences in their class. In alignment with our theoretical model, this procedure was designed to help students process their experiences as a means for improvement. 

\subsection{Prior Work: Iterations 1-3}

The first iteration of the GRF was developed in the context of a middle school science classroom. Starting from a set of valued character traits, two complementary rubrics were created: the \emph{Status and Progress Rubrics}. The Status Rubric allowed students to gauge their proficiency in 10 lifelong learning skills (e.g., courage, collaboration, and organization). The Progress Rubric, on the other hand, was meant as a tool for guiding students' development of skills in which they expressed interest in improving. Each rubric allowed students to rate themselves as \emph{beginning}, \emph{developing}, or \emph{succeeding} at skills or improvement plans.

The next iteration of the GRF was spearheaded by the Compass Project, a student-run organization whose mission is to improve equity in the UC Berkeley Physics Department through sociocultural support and other strategies~\cite{Albanna2013}. Compass adapted the Status and Progress Rubrics for use at the university level. In this process, Compass combined aspects of the Status and Progress Rubrics into a single rubric, now called the \emph{Guided Reflection Rubric}, which contains a similar set of skills found in the original rubrics. Students enrolled in Compass's fall and spring semester courses submitted weekly reflections based on the rubrics. Analyses of student reflections from one such Compass course indicated there were three major areas of focus: organization, connections, and persistence, which comprised 68\% of all reflections~\cite{gandhi_teaching}. 

The third iteration of the GRF involved creating a web form to streamline the reflection process. Given that most student reflections in the second iteration revolved around only a few themes (i.e., organization, connections, and persistence), these themes were made explicit in the online form. Students nevertheless had access to the full Guided Reflection Rubric, which was available electronically via a link from the online form. The online form was used used in multiple classrooms across the country. During Summer 2014, instructors who used the GRF during the previous academic year met at UC Berkeley to discuss and revise the online version of the form. This allowed experiences to be gathered from a variety of institutions, including: Arizona State University, Boise State University, Cal Poly San Luis Obispo, UC Berkeley, and the University of Maryland. These discussions were used as the basis for future work, with many of the ideas underlying incorporated to improve the nature of the prompts used in the fourth iteration, which is the subject of the present work.

\subsection{Current Design: Iteration 4}

The GRF required that students respond to seven prompts:

\begin{enumerate}
\item Think about a specific episode from last week that you would like to improve upon. Choose
the skills you think would help you improve on this in the future.
\begin{itemize}
\item Bouncing back from failure or other setbacks
\item Building a network and developing collaboration skills
\item Becoming an organized, self-aware, and mindful person
\item Something different.
\end{itemize}
\item Describe the specific experience from last week that you would like to improve upon.
\item What strategies did you use to respond to the challenge? (Check all that apply.)
\item Describe an aspect of this experience that you can improve in the future. (Provide at least one concrete strategy you will use to be more successful.)
\item What resources did you use this week? (Check all that apply.)
\item (Optional) Comment on your experience using these resources last week.
\item (Optional) Is there anything else that you would like to share?
\end{enumerate}

These seven prompts were designed: (1) to get students to revisit a salient experience from the previous week, (2) set a goal for improvement, and (3) and decide upon concrete steps for improvement. After students chose one of the four focal areas for reflection, the GRF presented a short paragraph describing the importance of these types of skills, which was designed to help students focus their recollection of a specific experience. The choice of area for reflection also influenced what students saw in the third prompt of the reflection. Depending on which area students focused on, they were given different specific strategies with checkboxes next to them. The purpose of steps (3) and (5), which focused on checkboxes of strategies and resources, was to remind students of potential resources and courses of action that were available to them on a regular basis. Other than this descriptive text and checkbox of strategies, all of the prompts were identical regardless of which area of focus students chose. If students chose to reflect on ``something different,'' they were asked to come up with their own list of strategies that they used.

After students completed the GRF, on a weekly basis, they received a typed response from their instructor. This response was emailed back to them in a packet that included their original reflection and the response so that they could be read together. The instructor focused on empathizing with the students and also gave concrete feedback and strategy suggestions about how students could improve. For instance, 
\begin{quote}
\emph{One strategy I use to help keep track of long-term projects is a big whiteboard. I break up my projects into smaller parts and assign deadlines for myself to get those pieces done. After I complete a task, I check it off but I don't erase it because it's important to me to be able to see the progress I've made towards the end product. Do you have or use a whiteboard? Do you think it will help you organize your long-term projects?} \student{Instructor}{3}
\end{quote}
In many cases the instructor commented not on the actual strategies that students suggested, but to what extent they were actionable and achievable:
\begin{quote}
\emph{\ldots time management skills are some of the most important skills I learned in college, and I can relate to being so busy with extracurricular obligations that reading assignments fall by the wayside. What are some concrete strategies you plan on using to manage your time better? What does ``planning ahead" mean to you?} \student{Instructor}{2}
\end{quote}
We do not discuss instructor feedback in detail, because our analyses showed that students rarely incorporated instructor feedback into their action plans, at least explicitly.

\section{Methods}

In this and subsequent sections, we follow the recommendations of Hammer and Berland~\cite{Hammer2014} when explicating our coding process and reporting data.

\subsection{Participants}

Participants in the present study were students enrolled in \emph{Teaching Physics}, an upper-division course for students considering a career teaching physics or other physical sciences. The course was based on the summer workshops used by the American Modeling Teachers Association to train physics teachers in the \emph{Modeling Instruction}~\cite{Brewe2008} approach to teaching physics. Such workshops typically span 100 hours over the course of three weeks during the summer. \emph{Teaching Physics} met twice per week for two hours per session, for a total of 40 hours over the course of a 10-week quarter. Accordingly, \emph{Teaching Physics} covered less content than a typical \emph{Modeling Instruction} workshop. Beyond introducing students to model-based physics pedagogy, \emph{Teaching Physics} further engaged students in two additional activities: discussions about the nature of intelligence, and Peer-Assisted-Reflection (PAR)~\cite{reinholz_assessment_2015}. Discussions about intelligence focused heavily on the dichotomy of growth- versus fixed- mindset~\cite{Dweck2006}. PAR activities complemented these discussions by giving students the opportunity to provide one another with feedback on homework assignments as well as to revise their own work based on input from their peers.

In total, 12 students were enrolled in \emph{Teaching Physics}, all of whom participated in the study. Five were Physics majors, four were majors in other STEM disciplines, and three were Liberal Studies majors (i.e., future K-8 teachers). The course was co-taught by two instructors, one of whom (D.R.D.F.) is an author of the present work. In this and subsequent sections, we present excerpts of reflections from all students throughout the course of the quarter. 

Seven participants were women, and five were men. The number of quotes from men and women presented here is representative of the demographics of the students enrolled in the course. However, to protect the identity of the students in this small sample, we do not identify students by their gender in this work.

\subsection{Data collection}

Students were required to submit weekly reflections using the GRF for the first nine weeks of the quarter. During the tenth and final week, students uploaded a final reflection that was not guided by the prompts found in the GRF. Six students completed all 9 weekly reflections, five completed 7 or 8, and one student only completed 5 weekly reflections. Out of 108 possible reflections, 97 were submitted, corresponding to an overall completion rate of 90\% for the weekly reflections. During the first 8 weeks of the quarter, between 10 and 12 students submitted their reflections each week. Completion rates dropped during the last two weeks of the quarter: in Week 9, only 6 of the 12 participants submitted their weekly reflections; and in Week 10, only 7 completed their final reflections.

Reflections were treated as additional homework assignments and were graded for completion. On eight of the weekly reflections, instructors provided students with personalized feedback, delivered in two formats: email and paper-copy. Electronic copies of students' reflections and corresponding instructor feedback were stored on a secure drive for research purposes. All data were collected electronically and excerpts are presented verbatim, except when redactions or other changes were necessary to protect participants' identities. No changes were made to correct for spelling, grammar, punctuation, or capitalization.

\subsection{Analysis}

Aligned with our model, our initial analyses focused on identifying the presence of three key aspects of reflection in their reflections: (1) narrating events, (2) making goal statements, and (3) making plans to meet those goals. Although the GRF requires students to respond to multiple prompts individually, we did not distinguish between parts of student reflections in analysis, instead treating each reflection as a whole unit. To analyze reflections, we developed a coding scheme over three iterations of design, coding, and establishing inter-rater reliability. To calibrate our coding, we collaboratively coded 26 reflections written by 3 students. The remaining 71 reflections (written by 9 students) were coded twice, once by each author independently. The final coding scheme that emerged from this process consisted of identifying the presence or absence of three characteristics for each student reflection: narrative, goal (growth and achievement), and action statements. To exemplify our coding scheme, we provide both examples and non-examples for each category. Here and henceforth, we denote students by anonymous identifiers S1 through S12.

\subsubsection{Narrative statements}

Narrative statements involve description of a single event or pattern of events, which is the first component of our model of reflection. We used the following operational definition when coding for the presence of narrative statements:
\begin{quote}
A narrative statement must refer to something that is ongoing or has already happened as opposed to something that may happen in the future. An event or pattern of events must have taken place; it is insufficient to describe an interpretation of an event or pattern of events (e.g., ``I realized I'm not on track"). Narrative statements typically uses past tense, though sometimes present tense is used to describe patterns of events (e.g., ``There is not much discussion in this group, another student just seems to take control").
\end{quote}
The following quote is an example of a narrative statement:
\begin{quote}
\emph{last week i had my first quiz for the year and i was very terrified because last quarter i failed every quiz midterm and final and so i really had to bounce back.} \student{S9}{2}
\end{quote}
This example was considered a narration of an event because it identified a specific event (having a quiz) which prompted reflection. On the other hand, consider the following non-example:
\begin{quote}
\emph{With lingering homework assignments, exams finally completed, and pending projects I was uneasy about the progress I had made and still need to make for the end of the quarter.} \student{S12}{8}
\end{quote}
This is not considered an example of narrating an event because the student was discussing their anxieties regarding things that were upcoming (deadlines), not events that had already taken place. Thus, it would be difficult for the student to reflect on how they acted in a situation as the situation has not yet arisen.

\subsubsection{Growth statements}

Growth statements relate to the second component of our model for reflection, setting a goal for improvement. Growth statements are aspirational, focused on one's desired state of proficiency at a given skill or task compared to their current level of proficiency. When coding for the presence of growth statements, we used to following operational definition:
\begin{quote}
A growth statement must use the phrases ``do more/less of," ``get better/worse at," ``improve upon," ``reinforce," ``strengthen," etc. There should be an explicit positive connotation. Growth statements typically use committal language in the future tense such as, ``I will" or ``I want to," though sometimes non-committal language is used (e.g., ``I should start being kinder to myself"). Statements that use non-committal language in the past tense cannot be used (e.g., ``I should have been kinder to myself").
\end{quote}
The following quote is a growth statement, describing how a student wants to get better at something (collaborating with others):
\begin{quote}
\emph{I would like to keep working towards growing with others in collaboration.} \student{S7}{5}
\end{quote}
Contrast this with the following non-example of a growth statement:
\begin{quote}
\emph{Over the weekend, I took a test for the credential program. \ldots I should have prepared a little bit more and taken practice tests from different sources to have a better idea of what the questions would be like.} \student{S4}{5}
\end{quote}
This is not a growth statement because the student is not using noncommittal language in the past tense (``I should have"). Here, the student does not describe what they would do differently in the future.

\subsubsection{Achievement statements}

Achievement statements are another form of setting goals for the future. An achievement statement is a description of what one wants to achieve or accomplish. These types of statements are indicative of a specific target for improvement, unlike growth statements which focus on general improvement along some spectrum. In practice, we found that achievement statements were extremely rare, and it was not possible to create an emergent operational definition for this statement type. Nevertheless, we present our initial definition of achievement statements:
\begin{quote}
An achievement statement uses phrases such as, ``I'm trying to get an A," ``I want to pass this exam," or ``I want to graduate this quarter."
\end{quote}
The following quote is an example:
\begin{quote}
\emph{I have recently developed a habit of taking my phone out in my classes the moment there is any sort of break in the lecture or activities. \ldots I know that this sort of habit is distracting and can be seen as disrespectful to my professors, so I definitely want to stop doing it.} \student{S5}{5}
\end{quote}
This is an example because the student has created a concrete, measurable objective, and it will be clear once they complete it. And a non-example:
\begin{quote}
\emph{During the first couple days of class, any group work we did, I automatically went to people I know in my major who I have had classes with before. I want to be more open to working with any student in the class.} \student{S1}{1}
\end{quote}
This is not an example because there is no specific target for improvement; while the student may become more open to working with other students in the class, there is no obvious measure of whether or not this goal has been achieved.

\subsubsection{Action statements}

Action statements describe what one will do differently in the future in order to attain a particular goal or change a pattern of behavior, which relates to the final component of our model for reflection. Although we often could not determine if a student actually followed through with their plans, if an individual does not articulate a specific course of action, it is unlikely that they will actually take concrete steps towards improvement. In our coding scheme, we used the following operational definition to identify action statements:
\begin{quote}
An action statement must describe at least one instance of an activity in which the student intends to engage in the next week. The activity must be in response to a circumstance or as part of meeting a goal. The statement must include logistical information, such as when and/or where the activity will happen. The reader must be able to envision what the student will be doing.
\end{quote}
The following quote is an example of an action statement:
\begin{quote}
\emph{I printed out a calendar and will be marking all my assignments and when they are due.} \student{S5}{2}
\end{quote}
This example describes concrete actions that one could envision the student engaging in next week. Consider the following non-example:
\begin{quote}
\emph{My strategy for improving myself in this way is to be conscious of any time that I decide not to do an action because it may be too hard for me or make myself uncomfortable. When this happens, I will remember about the growth mindset and do the action anyways.} \student{S10}{2}
\end{quote}
This is not an action statement because it is unclear exactly when the student would engage in this action (when he feels uncomfortable). Also, there is no guarantee that such a situation would arise in the next week or during the semester at all.

\subsection{Categorizing Components}

Once we identified the various components of reflections, we double coded the components along various dimensions. Narrative events were coded as to whether or not they focused on: (1) in-class events, (2) out of class events, (3) exams, (4) other academic issues, or (5) non-academic issues. Reflections fit cleanly into only one of these categories in each case. Regarding growth and action statements, we coded whether they focused on (1) organization, (2) collaboration, (3) perseverance, or (4) other skills. Because achievement statements were so rare, they were dropped from our analyses.

\section{Results and Discussion}

We organize our results and discussion into four parts: (A) presence and frequency of narrative, growth, achievement, and action statements; (B) student use of multiple statement types during reflection; (C) students' reflections on the reflection process at the end of the semester; and, (D) description of the content of student reflections. Parts (A)-(C) relate to our first research question, describing how the GRF supported reflection, and part (D) relates to our second research question, describing the areas of focus for reflections. In each part, we distinguish between reporting data and interpreting results by using headers to signify interpretation.

\subsection{Presence and Frequency of Statement Types}

\tabIRR

To determine the inter-rater reliability of our coding scheme, we computed both the percent agreement and Cohen's unweighted kappa statistic~\cite{Cohen1960} for the 71 reflections coded independently by the authors. We conclude that there was substantial to almost perfect agreement on narrative, growth, and action statements~\cite{Blackman2000}; results are summarized in Table~\ref{tab:IRR}. For these statements, discrepant  codes were resolved through discussion. In total, we identified 77 narrative statements, 50 growth statements, and 38 action statements across all 97 reflections.

While we identified many examples of narrative, growth, and action statements in the calibration data set, neither rater identified any achievement statements during the calibration process. Lack of calibration on achievement statements lead to relatively low (i.e., moderate) agreement on this statement type~(Table~\ref{tab:IRR}). Moreover, both raters agreed that there were no achievement statements in 62 (87\%) of the 71 reflections that were coded separately. Given the low frequency of, and low agreement on, achievement statements, we did not attempt to resolve discrepant codes. We omit these statement types from further analyses.

\figHIST

Almost all (94\%) of the 97 reflections contained at least one narrative, growth, or action statement. Moreover, by the end of the quarter, 11 of the 12 students incorporated at least one of each type of statement in their reflections; the remaining student incorporated both narrative and action statements, but no growth statements. Reflections that included neither narrative, growth, nor action statements were neither common nor an artifact of any one particular student: the 6 such reflections in our dataset were written by 5 unique students. Fig.~\ref{fig:hist} shows the frequency with which students made each type of statement (or no type of statement) in their reflections. On average, students included narrative, growth, and action statements in about 80\%, 50\%, and 40\% of their reflections, respectively.

\subsubsection*{Interpretation}

Our results indicate that the GRF supported students to articulate the crucial components of reflection, as according to our model. This demonstrates the value of the GRF, as simply asking students to reflect may not result in reflection at all~\cite{anson_reflection_1997}. The high frequency of narrative statements is an expected outcome of the design of the online form, which asks students to ``describe a specific instance from last week." A lower rate of growth statements can be understood as a consequence of the fact that aspirations for improvement likely are not realized on a timescale of one week and therefore students may not feel the need to re-articulate the same growth statement over and over.

The lower rate of action statements could be a result of our prompts, which ask students to ``describe an aspect of this experience that [they] can improve in the future," rather than asking students to outline a plan for improvement. Despite the lack of emphasis on actionable plans in the tool itself, the idea that students should be making plans was reinforced by the instructors through their weekly feedback.

The reason for the dearth of achievement statements is unclear. Because the course included discussions about growth mindset, it could be that students were primed to think of achievement-based (extrinsic) motivation as inferior to growth-oriented (intrinsic) motivation, or to intuit that their instructors felt this way. Whether achievement-based goals would be articulated in different student populations or under different conditions is an open question. 

\subsection{Use of Multiple Statement Types During Reflection}

About two thirds of reflections contained more than one type of statement, usually consisting of a narrative statement coupled with growth, action, or both growth and action statements. Fig.~\ref{fig:venn} shows a breakdown of reflections according to statement types. The following example demonstrates use of narrative, growth, and action statements in a single reflection:
\begin{quote}
\emph{Last week I got my ass royally whooped by a programming assignment\ldots I want to improve on becoming a better programmer, and I think that that hinges on my being more organized. For this week, I will make sure to a lot at least 1 hour per day on the assignment due Wednesday, and 1 hour per day after that on the assignment due the following Wednesday\ldots I do not want to be staying up until midnight working on assignments anymore, so I will try to get everything done as early as possible.} \student{S10}{4}
\end{quote}
This quote begins with the student \textit{narrating} their experience of struggling with a challenging assignment for a programming class. Building on this, S10 expresses a desire for \textit{growth} in improving their organizational skills (in service of improving his programming skills), and creates an \textit{action} plan which consists of starting to work on assignments two weeks in advance the due date for an hour per assignment per day.

\figVENN

About a third of reflections contained only one type of statement, most often a narrative statement. For example:
\begin{quote}
\emph{This past week I had some trouble with the par packet problems. I was stumped and needed some guidance. I looked some parts up online but what helped most was meeting up with my peers. They explained it to me and that's what really helped me. I was happy that I went out and found help when I needed it and tried many approaches to solve the problem.} \student{S7}{3}
\end{quote}
Here S7 narrates their experience of struggling with the PAR problem and getting help from peers, but does not set any targets for growth or actions to achieve those targets.

Only 6 reflections contained neither narrative, growth, nor action statements. Towards the end of the quarter, {S4} submitted the following reflection:
\begin{quote}
\emph{My last week has been quite good, free from stress, and I honestly cannot think of anything that sticks out that I would need to consciously make a note to improve on.} \student{S4}{7}
\end{quote}
In this reflection, {S4} is not narrating a specific experience from the previous week. Neither are they identifying a goal for personal growth or an action plan for practicing a skill; indeed, {S4} explicitly states that they cannot think of anything to improve upon.

\subsubsection*{Interpretation}

Given that a majority of reflections contained a combination of narrative, growth, and/or action statements, we conclude that the GRF supported students in structuring their reflections in productive ways. Individual students interacted with the GRF in different ways; for example, while S11 included two or three statement types in almost all of their reflections, S2's reflections often included only narrative statements. Nevertheless, the tool was successful in engaging all students in multiple aspects of reflection.

The rare cases in which student reflections contained none of the statement types can potentially be explained by the framing of the activity; the GRF prompts students to think about a specific episode they would like to improve upon. As a result, during times when a student is feeling on top of things, the prompts might not resonate with them.

\subsection{Final Reflections}
Of the 7 students who completed the final reflection, 3 discussed the weekly reflection assignment despite the lack of an explicit prompt to do so. In each of these 3 cases, the students described favorable experiences with reflection. For example:
\begin{quote}
\emph{by using the GRF through out the quarter i was able to describe my feelings and emotion sand become more aware of my stress levels. this helped me be manage my stress more because i was actually aware of how i was feeling.} \student{S9}{10}
\end{quote}
Here {S9} is describes how the GRF not only helped them become more aware of how they were feeling, but also helped them to manage their stress.

While {S9} spoke to development of a particular skill--in this case, mindfulness--other students comment on the practice of reflection more generally. For instance:
\begin{quote}
\emph{Setting concrete and obtainable goals was also a practice I learned and continue to work towards\ldots there were many areas that I feel I have grown as a physics students/teacher during this Spring quarter. The most important to me and always applicable overarching theme of loving to learn.} \student{S12}{10}
\end{quote}
Here {S12} is not referring to a particular skill like time management or teamwork, but instead to the practice of setting goals--an important aspect of the reflection process. {S12} sees this practice as connected to their growth in many areas as well as their love of learning.

A different student, {S10}, articulates similar appreciation of the abstract process of reflection:
\begin{quote}
\emph{These reflections have been surprisingly significant in my life, and have really helped me identify problems clearly and come up with solutions, and feedback from you, the teachers, has really helped. I used to journal a lot in high school and haven't for a long time, this class has inspired me to take some time every week and reflect, and I plan on continuing that practice after the class is over.} \student{S10}{10}
\end{quote}
{S10} views that GRF as a significant tool for identifying and overcoming challenges. While {S10} acknowledges the role of instructor feedback in this process, the GRF activity has inspired them to start reflecting on their own. Because {S10} invokes their past experience with journaling, it is likely that his vision for continued reflection does not require continued feedback from instructors.

\subsubsection*{Interpretation}

Though only a few students wrote explicitly about the GRF in their final reflections, those who did so expressed overall positive experiences with the activity. Students spoke in abstract terms about the usefulness of reflection in identifying problems, setting goals, and overcoming challenges. This evidence suggests that the GRF was successful in supporting productive student reflection.

\subsection{Content of reflections}

The content of reflections is summarized in Table~\ref{tab:narrative-themes}. About half of the 77 narrative statements referred to out-of-class activities, including weekly homework assignments, studying, and long-term projects. For example, {S1} describes a pattern of working on homework assignments near their due dates:
\begin{quote}
\emph{I end up doing some homework assignments right before they are due regardless of when they were assigned.} \student{S1}{4}
\end{quote}

\tabNARR

About a quarter of narrative statements referred to in-class activities, such as working in groups, completing class activities, and giving presentations. For example, {S8} describes an in-class presentation:
\begin{quote}
\emph{I made a mistake during presentation of an explanation for a in class exercise.} \student{S8}{3}
\end{quote}

The remaining quarter of narrative statements referred to other academic and non-academic episodes. For example, {S1} describes their experience meeting the requirements for completing their degree:
\begin{quote}
\emph{Having the only 4 classes I have left as an undergrad be messed up in the system threw me for a loop I was not expecting or prepared for.} \student{S1}{7}
\end{quote}
In this quote, {S1} is narrating an \emph{academic episode}--namely, changes to their schedule which may affect timely graduation--which is unrelated to things like doing homework, working in groups, etc. The following quote, on the other hand, provides an example of a \emph{non-academic} narrative statement, in which {S4} articulates frustration with the cancelation of a sports meet.
\begin{quote}
\emph{I just found out today that the meet that I signed up for did not get enough people for it to span over two days as originally planned\ldots I am now unable to attend and it is very frustrating that I have been preparing so specifically for this meet and will not be able to compete.} \student{S4}{4}
\end{quote}

Whereas narrative statements refer to episodes that can be described thematically, growth and action statements are more aligned with skills which students would like to improve, practice, or employ. Accordingly, we collaboratively coded growth and action statements according to the skills outlined in the online form which students used to submit their reflections: \emph{organization}, \emph{collaboration}, \emph{resilience}, and \emph{other}. When submitting a reflection via the online form, students were required to identify the skill upon which their reflection would focus. We compared students' selection of skill on the online form to our skill codes as a means of investigating the fidelity of this feature of the online form. We found that students' skill selection is aligned with our codes in 36 of 50 (72\%) of growth statements and 27 of 38 (71\%) of action statements. A breakdown of growth and action statements by \emph{coded} skill type (as opposed to student selection of skill type via the online form) is given in Fig.~\ref{fig:pie}.

\figPIE

About half of growth statements and over two thirds of action statements focused on organizational skills, most often time management. Growth and action statements focused on developing time management skills were made primarily in three contexts: about a third were made in conjunction with narratives about out-of-class homework; another third were made in conjunction with narratives that spanned a broad set of academic and non-academic experiences; and the final third were made in the absence of any narrative statement at all.

Almost all narratives about homework were accompanied with a growth or action statement about time management. Consider the following reflection:
\begin{quote}
\emph{I did a few of the homework problems ahead of time, and then procrastinated to do the rest until the night before; only to find that the last problems were very involved and difficult. This resulted in me not finishing some of the problems\ldots Making sure I attempt every problem at least 2 or 3 nights before the problem set is due would be extremely helpful and would allow me to visit [my professor's] office hours the day before it is due.} \student{S2}{3}
\end{quote}
In this reflection, {S2} narrates an experience about homework: they failed to allot enough time to complete a challenging assignment on time. S2 accompanies their narrative statement with an action plan about time management: {S2} will start working on their assignments earlier in the week, giving them enough time to attend the only available office hour which fits with their schedule.

Similarly, most reflections that did not include any narrative statement included a growth or action statement about time management. For example, one student developed a plan for dealing with a large course project (a future event), so it was not coded as narrating an event that already happened:
\begin{quote}
\emph{This week I do not have a very large assignment due, but have been assigned a large project that is due near the end of the quarter. So, this week I want to work on things even though they are not due for a long time. More specifically, I will create an outline for my programming assignment by this weekend, and be capable of presenting it before arranging a meeting with other students who are working on a similar project.} \student{DL}{5}
\end{quote}

About a quarter of both growth and action statements focused on collaboration skills, including teamwork and networking. Almost all growth and action statements focused on teamwork skills were made in conjunction with a narrative about in-class group work. Similarly, whenever students narrated an experience that had to do with in-class group work, they almost always made a growth or action statement focused on teamwork skills. For example, {S2} articulates a desire to grow confident and contribute more to their group:
\begin{quote}
\emph{I've found that in my new group \ldots it's more difficult to me to find my voice and contribute to what we're doing. There is not much discussion in this group, [another student] just seems to take control \ldots I'd like to be more confident in my group, and contribute more.} \student{S2}{4}
\end{quote}
In the next example, S7 narrates a positive experience working with not only a new student they hadn't previously met, but a student from a group (men) with whom they were unused to working. In addition to sharing this positive experience, S7 sets an action goal of asking more questions of the quieter person in the group:
\begin{quote}
\emph{I was nervous this last week to work with a new person that I have never met before. I have had many classes with girls so working with them is what I am used to. This week I worked with [a male student] for my par packet and in the pendulum problem. It ended up being really good because we worked well as a big group together. I think next time I can make sure that being in a big group I ask questions of the quieter person so they can be more of a part of the team.} \student{S7}{1}
\end{quote}

Resilience skills were relatively uncommon among reflections. No action statements involved resilience skills and only 10\% of growth statements did so. Most of these reflections focused on self-compassion rather than perseverance or intellectual courage. For example:
\begin{quote}
\emph{From the beginning of the course I felt misplaced in this course\ldots However after being able to work with the other students I feel more welcomed than before\ldots During this week for [PAR] my "grader" was really interested in the way I responded to the activity. Two of my classmates also asked me if I could clarify them some things from [PAR]. I was more than happy to help them. I think I should not be so harsh on myself in thinking that I do not have to knowledge or experience to be in this course. Just to keep trying, working hard, and asking more questions. Yes the feeling of not being part of the class because everyone else is in some sort of teaching path and I going into [a different field] is going away. I feel more welcomed and not out of place.} \student{S8}{4}
\end{quote}
Here {S8} articulates a desire to grow their self-compassion skills, by being less harsh on themselves for feeling misplaced (in this case, due to perceived lack of relevant content knowledge, experience, and intended career path). 

Finally, 18\% of growth and 8\% of action statements focused on skills that could not be binned into these categories. For example, many of these statements focused on communication skills:
\begin{quote}
\emph{I think giving presentations on the fly is something I need to work on being more confident with. I'm usually fine with public speaking, but sometimes when I wing it I feel a bit awkward and rambly.} \student{S2}{7}
\end{quote}
The other reflections were mostly idiosyncratic, such as the following, focused on labs:
\begin{quote}
\emph{I'd like to take the labs more seriously and try to empathize more with what a student may be going through when trying to read and understand data they take.} \student{S2}{6}
\end{quote}

\subsubsection*{Interpretation}

The breadth of narrative themes is consistent with the framing of the reflection assignment, in which students were tasked with reflecting on any aspect of their college experience. Difficulties related to time management with homework and studying seem to be common to students of all ages across institutions~\cite{gandhi_teaching, zaniewski_worries}. However, given the nature of the \emph{Modeling Instruction} pedagogical approach, in-class themes that focus on group work and presentations may be more prevalent in this context than in others where different approaches to teaching and learning are employed. Such themes may also be prevalent in project-based learning contexts~\cite{ward_role_2014}.

Non-academic themes can reveal emotional or financial hardship, which can be calls for intervention or opportunities for interpersonal bonding. The frequency of such narratives likely depends on the nature of the student-teacher relationship being cultivated. Non-academic themes may also be mostly unproductive, such as when students simply focus on activities such as weight lifting.

The lack of resilience-oriented action statements is noteworthy given the course emphasis on growth mindset. Also, we found that persistence was a major theme for first-year students at Berkeley~\cite{gandhi_teaching}, but it was not prevalent in the present study. It may be that students in their junior year already have functional coping mechanisms since most attrition happens in the first or second year of college. Such statements may also be infrequent due to the dearth of tools that support resilience (e.g., in contrast to time management).

\section{Summary and Future directions}

This paper addressed two main research questions: (1) how did the GRF support student reflections? and (2) what were the specific areas of focus for student reflections? Addressing the first question, the GRF appeared to support reflection in meaningful ways. Students frequently engaged in all three components of reflection~\cite{boud_promoting_1996}, and  were often able to coordinate these components together in useful ways. In their end of semester reflections, some students also noted the importance of reflection and how the GRF supported them through the process.

Our analyses also provided insight into the actual themes present in student reflections. The majority of reflections were related to academics, with about half of all reflections focused on out of class activities. One of the most common themes was time management, which appears to be an issue across physics contexts~\cite{gandhi_teaching, zaniewski_worries}. Given the prevalence of such reflections, instructors may benefit from addressing this explicitly in their courses. It was also relatively difficult for students to develop growth statements and action plans related to collaboration and persistence. Even though these were important topics, students may require greater scaffolding and support to reflect on them meaningfully. 

\subsection{Open questions}

Some aspects of student reflections were not captured by our scheme, such as personal anecdotes that provided insight into the students' lives and deepened their bond with their instructor. For example:
\begin{quote}
\emph{there is this girl at church that i really like and thats cool cause i got to talk to her this weekend.} \student{S9}{4}
\end{quote}
Students also used the GRF for in-depth processing of experiences that was not necessarily related to the development of action plans:
\begin{quote}
\emph{It seems I have three types of procrastination: Productive, where I do something else I find more important or interesting than the assigned task; Distracting, where I fill the time with pointless activities like internet surfing; and Paralyzing, where some part of me is so set against doing the task at hand that I instead do absolutely nothing (but fidget and think) for hours\ldots It's the paralysis that I want to put an end to. Unfortunately, it's a problem I've had since elementary school, and I still don't know what to do about it.} \student{S11}{6}
\end{quote}
In this reflection, the student is engaged with multiple complex issues, like identity as a physicist versus physics teacher, motivation and procrastination, etc. Our scheme does not capture this complexity.

\subsection{Implications for design}

Although the GRF supported students to reflect on specific experiences, students' action plans were less frequent than desired. In the future, the reflection prompts will be revised to explicitly ask students to make action plans. Also, the online format made it possible for students to complete weekly reflections without looking over the feedback they received or the goals and plans they set for themselves in previous weeks. Future versions should create opportunities for students to re-read their past reflections and make sure they are working towards the same goal for multiple weeks, thus creating opportunities to discuss revisions and refinements of plans.

Finally, peer and instructor feedback were key features in this environment, a fact which must be taken into consideration in the process of refining these tools or adapting them for use in other settings. Feedback was important to students. For example:
\begin{quote}
\emph{The biggest single thing I took away was the development of the notion of effective feedback. As far as course materials, these ideas were introduced in readings, and the other activities served as practice for implementing those ideas.} \student{S11}{10}
\end{quote}
Here it is likely that {S11} is referring to their ability to give effective feedback on PAR problems. In the future, a revised version of the GRF that allows students to provide feedback to one another may create interesting new learning possibilities, as students often learn as much from giving feedback as getting it~\cite{reinholz_assessment_2015}. This is a subject of ongoing design and research.

\acknowledgments The authors acknowledge Chance Hoellwarth and Danielle Champney for their contributions to the design of this study, and Angela Little and Bethany Wilcox for guidance on the communication of results. This work builds off of discussions from the 2014 Prism Summer Retreat, which was funded by the AAPT Physics Education Research Leadership and Organizing Council and the Center for Excellence in STEM Education at Cal Poly San Luis Obispo. In addition, D.R.D.F. was supported by NSF grant DUE-1323101, and D.L.R. by the AAU Undergraduate STEM Education Initiative.

D.R.D.F. and D.L.R. contributed equally to this work.


%

\end{document}